\documentclass[useAMS,usenatbib]{mnras}  
\usepackage{graphicx}
\usepackage{color}

\title[Origin of the soft photon of PKS 1424+240]{On the origin of the soft photons of the high synchrotron peaked blazar : PKS 1424+240}
\author[S. J. Kang et al.]
{Shi-Ju Kang$^{1,2}$\thanks{E-mail:~kangshiju@hust.edu.cn},
Yong-Gang Zheng$^{3}$,
Qingwen Wu$^{2}$\footnotemark[1]\thanks{Corresponding author: qwwu@hust.edu.cn}
and Liang Chen$^{4}$\\
$^{1}${Department of Physics and Electronics Science, Liupanshui Normal University, Liupanshui, Guizhou, 553004, China} \\
$^{2}${School of Physics, Huazhong University of Science and Technology, Wuhan, Hubei, 430074, China}\\
$^{3}${{Department of Physics, Yunnan Normal University, Kunming, Yunnan, 650092, China}}\\
$^{4}${Key Laboratory for Research in Galaxies and Cosmology, Shanghai Astronomical Observatory} \\
{~~Chinese Academy of Sciences, 80 Nandan Road, Shanghai, 200030, China}\\
}

\begin{document}

\date{Accepted Year Month Day. Received Year Month Day; in original form Year Month Day}

\pagerange{\pageref{firstpage}--\pageref{lastpage}} \pubyear{2016}

\maketitle

\label{firstpage}

\begin{abstract}
  PKS 1424+240 is a distant very high energy gamma-ray BL Lac object with redshift $z=0.601$. It was found that pure synchrotron
  self-Compton (SSC) process normally need extreme input parameters (e.g., very low magnetic field intensity and extraordinarily large
  Doppler factor) to explain its multi-wavelength spectral energy distributions (SEDs). To avoid the extreme model parameters,
  different models have been proposed (e.g., two-zone SSC model or lepto-hadronic model). In this work, we employ the traditional
  one-zone leptonic model after including a weak external Compton component to re-explore the simultaneous multi-wavelength SEDs of
  PKS 1424+240 in both high (2009) and low (2013) states. We find that the input parameters of magnetic field and Doppler factor are
  roughly consistent with those of other BL Lacs if a weak external photon field from either broad line region (BLR) or the dust torus.
  However, the required energy density of
  seed photons from BLR or torus is about 3 orders of magnitude less than that constrained in luminous quasars (e.g., flat-spectrum
  radio quasars, FSRQs). This result suggests that the BLR/torus in BL Lacs is much weaker than that of luminous FSRQs (but not fully
  disappear), and the inverse-Compton of external photons from BLR/torus may still play a role even in high synchrotron peaked blazars.
\end{abstract}

\begin{keywords}
{BL Lacertae objects: individual (PKS 1424+240, VER J1427+23)}
\end{keywords}

\section{Introduction}

    Blazars, including flat-spectrum radio quasars (FSRQs) and BL Lacertae objects (BL Lacs), are radio-loud active galactic nuclei (AGNs)
    with a relativistic jet pointed at a small viewing angle to the line of sight. The multi-wavelength spectral energy distributions (SEDs)
    from the radio to the $\gamma$-ray bands of blazars dominantly come from the non-thermal emission, where the SED normally exhibits a
    two-hump structure in the $\nu-\nu F_{\nu}$ space. The lower energy hump is normally attributed to the synchrotron emission produced
    by the non-thermal electrons in the jet while the second hump mainly come from inverse Compton (IC) scattering. The seed photons for
    IC scattering may come from the synchrotron photons (SSC process, e.g., \citealt{1981ApJ...243..700K}; \citealt{1985ApJ...298..114M};
    \citealt{1989ApJ...340..181G}) and/or external photons (EC process), where the external photons possibly originate from the accretion
    disk (e.g.,\citealt{1993ApJ...416..458D}; \citealt{1997A&A...324..395B}), the broad line region (BLR; e.g., \citealt{1994ApJ...421..153S};
     \citealt{1996MNRAS.280...67G}), and/or the molecular torus (e.g., \citealt{2000ApJ...545..107B}; \citealt{2008MNRAS.387.1669G}).

   PKS 1424+240 (VER J1427+237) {is possibly the most distant BL Lac object (or the third most distant blazar) with a redshift of
   $z=0.6010\pm0.003$ \citep{2016A&A...589A..92R} that detected the very high energy (VHE; $E\geq 100$ GeV) $\gamma$-ray emission up to now. The other two VHE blazars are FSRQs (S3 0218+357, $z=0.944\pm0.002$, \citealt{2003ApJ...583...67C}, \citealt{2014ATel.6349....1M}, \citealt{2015arXiv150804580S} and PKS 1441+25, $z=0.9397\pm0.0003$, \citealt{2012ApJ...748...49S,2015ATel.7416....1M})}. PKS 1424+240 was first observed in the radio band
   (408 MHz) in {the} 1970s \citep{1974A&AS...18..147F}, {and} was categorized as a blazar based on its optical
   polarization (\citealt{1988ApJ...333..666I} and \citealt{1993AJ....106.1729F}). Based on its weak emission-line feature and the peak
   frequency of the lower energy bump,
   PKS 1424+240 is categorized as an intermediate-energy-peaked BL Lac (IBL) (\citealt{2006A&A...445..441N}), or a high-energy-peaked BL Lac (HBL) (\citealt{1996MNRAS.279..526P}). PKS 1424+240 was most recently classified as a high synchrotron peaked (HSP, which is the commonly utilized categorization now; \citealt{2010ApJ...716...30A}) BL Lac object in \cite{2014arXiv1403.4308A}.

   PKS 1424+240 was detected in
   $\gamma$-rays by the Fermi Large Area Telescope (LAT; \citealt{2009ApJ...697.1071A}) with a very hard spectrum (photon index $\Gamma = 1.85 \pm 0.07$,
    \citealt{2009ApJ...707.1310A}). The VHE $\gamma$-ray emission was also detected in Spring of 2009 by VERITAS \citep{2009ATel.2084....1O} and
    confirmed as a VHE emitter by MAGIC \citep{2009ATel.2098....1T}. Different models have been proposed to explain its hard VHE $\gamma$-ray spectrum.
    One-zone SSC model normally require a very large Doppler facotor of $\delta\sim50-130$ and a quite weak magnetic field of $\sim 0.02$ G (e.g.,
    \citealt{2010ApJ...708L.100A}, \citealt{2014A&A...567.135A}). To overcome these extreme parameters, \cite{2014A&A...567.135A} proposed a
    two-zone SSC model and found that it can roughly describe multi-wavelength SED of PKS 1424+240 with reasonable model parameters. \cite{2015MNRAS.447.2810Y} suggested that
    the extreme model parameters can also be avoided if adopting a lepto-hadronic hybrid jet model, where the VHE {$\gamma$-ray emission}
    is attributed to the synchrotron emission of pair cascades resulting from $p\gamma$ interaction (\citealt{2015MNRAS.449.1018Y}).

   Some recent works found that it can overcome the extreme parameters in a few BL Lacs if including a weak EC component (\citealt{2008ApJ...684L..73A};
   \citealt{2011ApJ...726...43A}; \citealt{2012AIPC.1505..647P}; \citealt{2013arXiv1312.4829L}; \citealt{2014ApJ...783...83L}). To understand the VHE
   emission in the most distant {BL Lac} of PKS 1424+240, we try to explore whether the traditional one-zone leptonic model after including a weak
   EC component can explain its multi-wavelength SED or not. Throughout the letter, we assume the following cosmology: $H_{0}=70\ \rm km\ s^{-1} Mpc^{-1}$,
   $\Omega_{0}=0.3$ and $\Omega_{\Lambda}=0.7$.

\section{The Model}{\label{sec2}}
  In this work, we adopt the traditional one-zone synchrotron + IC model to fit the SEDs of PKS 1424+240, which is widely used in blazars
  \citep[see e.g.,][and references therein]{2010MNRAS.402..497G}. A homogeneous sphere with radius $R$ embedded in a magnetic field $B$ is assumed,
  that moves relativistically with a speed of $\upsilon=\beta c$ ({bulk Lorentz factor} $\Gamma=1/\sqrt{1-\beta^2}$) along the jet orientation. Doppler factor $\delta=\left[\Gamma\left(1-\beta\cos\theta\right)\right]^{-1}\approx\Gamma$
  is assumed for the relativistic jet with a small viewing angle $\theta\leq 1/\Gamma$. The electron spectrum is assumed as a broken power-law distribution,
  with {indices} $p_{1}$ and $p_{2}$ below and above the break energy $\gamma_{b}m_{e}c^{2}$,
   \begin{equation}
   N(\gamma )=\left\{ \begin{array}{ll}
                    N_{0}\gamma ^{-p_1}  &  \mbox{ $\gamma_{\rm min}\leq \gamma \leq \gamma_{\rm b}$} \\
            N_{0}\gamma _{\rm b}^{p_2-p_1} \gamma ^{-p_2}  &  \mbox{ $\gamma _{\rm b}<\gamma\leq\gamma_{\rm max}$}
           \end{array}
       \right.
  \label{Ngamma}
  \end{equation}
  where $\gamma_{\rm min}$ and $\gamma_{\rm max}$ are the minimum and maximum electron Lorentz factors, and $N_{0}$ is the normalization of the particle
  distribution. Such a broken power-law distribution is a steady-state electron spectrum, which could be the result of the balance between the particle
  cooling and escape rates in the blob (e.g., \citealt{1962SvA.....6..317K}; \citealt{1994ApJ...421..153S}; \citealt{1996ApJ...463..555I};
  \citealt{1998A&A...333..452K}; \citealt{1998MNRAS.301..451G}; \citealt{2002ApJ...581..127B}; \citealt{2012ApJ...748..119C}; \citealt{2013ApJ...768...54B}).

  Different from the FSRQs, we consider a possible weaker external Compton process (e.g., external photons from BLR or torus) in this HSP BL Lac, where
  the external seed photon field may still play some roles in Compton process in some BL Lacs (e.g., \citealt{2002ApJ...581..143B,2002ApJ...581..127B}).
  Since the location of the $\gamma$-ray emission region is still unclear, we assume the external seed photons predominantly originate either from the BLR or from the dust torus.
  The external radiation field is characterized by an isotropic blackbody with the temperature
  $T=h\nu_{\rm p}/(3.93k_B)$, where $\nu_{\rm p}$ is the peak frequency of seed photons in the $\nu-\nu F_{\nu}$ space. For the BLR cloud, the most
  prominent contribution comes from the Ly$\alpha$ line, and hence the spectrum is assumed to be a blackbody with a peak around $2\times10^{15}~\Gamma$
  Hz \citep[see,][]{2008MNRAS.387.1669G}. For the IR torus, the spectrum is assumed to be a blackbody with a peak frequency of
  $\nu_{\rm IR} = 3 \times 10^{13}~\Gamma$ Hz in the comoving frame \citep{2007ApJ...660..117C}. The Klein-Nishina effect in the inverse Compton scattering
  and the self-absorption effect in synchrotron emission are properly considered \citep[see,][]{1979rpa..book.....R, 1970RvMP...42..237B}.

  PKS 1424+240 is a distant VHE blazar and the VHE emission is expected to be significantly absorbed by the extragalactic background light (EBL) via pair production. The absorption of gamma-rays by the EBL can be estimated using the model-dependent gamma-ray opacity of $\tau(\nu,z)$, where the relation between the observed spectrum, ${F}_{\rm obs}(\nu)$, and the intrinsic spectrum, ${F}_{\rm in}(\nu)$, can be described by the following relation,
  \begin{equation}
  F_{\rm obs}(\nu,z) = e^{-\tau(\nu,z)}F_{\rm in}(\nu,z),
  \end{equation}
  where $\tau(\nu,z)$ is the absorption optical depth due to interactions with the EBL (\citealt{{2004A&A...413..807K}}; \citealt{2005ApJ...618..657D}; \citealt{2012MNRAS.422.3189G}; \citealt{2008A&A...487..837F}; \citealt{2010ApJ...712..238F}; \citealt{2010A&A...515A..19K}; \citealt{2011MNRAS.410.2556D}; \citealt{2012Sci...338.1190A}; \citealt{2013A&A...550A...4H}).
  In order to minimize hardening introduced from EBL absorption corrections, we adopt the absorption optical depth that derived from the EBL model proposed by \cite{2011MNRAS.410.2556D} in our calculations. In our SED modeling of Figure~\ref{fig:2009p2013}, we assume the model prediction as the intrinsic emission and correct it to our local universe using equation (2), and compare it with the observational data (e.g., \citealt{2011ApJ...728..105Z}; \citealt{2013ApJ...764..113Z}; \citealt{2013MNRAS.431.2356Z,2014MNRAS.442.3166Z}; \citealt{2014ApJS..215....5K}).

\section{{Modeling the SEDs of PKS 1424+240}}{\label{sec3}}

\begin{figure*}
\centering
\includegraphics[width=16cm,height=15cm]{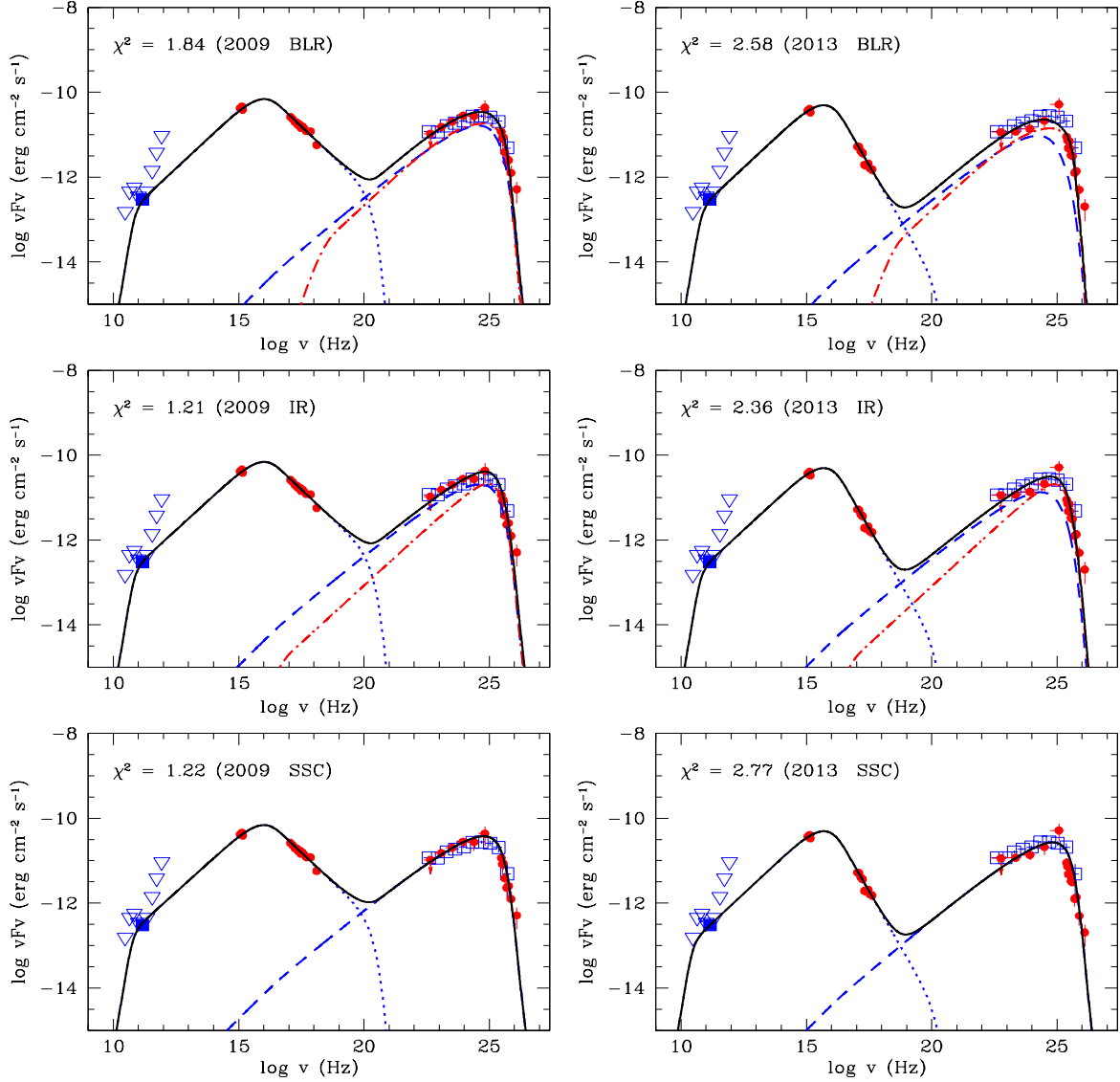}
\caption{The SEDs of PKS 1424+240. The red solid points represent the broadband simultaneous observational data,
empty blue squares indicate extended LAT data set (MJD 54682 to 56452), the non-simultaneous radio to submm data from Plank (30 GHz to 857 GHz)
are also plotted (blue triangles represent the upper limit, see \citep{2012A&A...541A.160G}. The dotted, dashed, dot-dashed and solid lines
represent the synchrotron, SSC, EC and total emission respectively. The left column and right column show the SEDs of high state in 2009 and
low state in 2013 respectively, while the upper, middle, and lower panels correspond to the fitting result using different models
(EC-BLR, EC-IR and SSC).}
\label{fig:2009p2013}
\end{figure*}

   The simultaneous multi-wavelength data from  VHE (VERITAS), HE ($Fermi$) $\gamma$-ray, X-ray and UV ($Swift$) for PKS 1424+240 at two different
   states are collected, where the high state in 2009 and low state in 2013 have integral flux above 120 GeV around
   $(2.1\pm0.3)\times10^{-7} {\rm ph}~{\rm m}^{-2}~{\rm s}^{-1}$ and $(1.02\pm0.08)\times10^{-7} {\rm ph}~{\rm m}^{-2}~{\rm s}^{-1}$,
   respectively (see \citealt{2014arXiv1403.4308A} for more details on the multi-waveband data). In Figure~\ref{fig:2009p2013}, the red solid
   points represent the simultaneous observational data (in 2009 and in 2013), empty blue squares indicate the extended LAT data set (MJD 54682
   to 56452; see~\citealt{2014arXiv1403.4308A}). The non-simultaneous radio to sub-millimeter data (30 GHz to 857 GHz) from Plank were also plotted (blue
   solid square, the down triangles represent the upper limits, \citealt{2012A&A...541A.160G}).

  We apply the one-zone jet model as described in Section 2 to reproduce the multi-waveband SEDs of PKS 1424+240. There are 9 {and} 10
  parameters in the pure SSC {and} SSC+EC models respectively: $R$, $\delta$, $B$, $p_{1}$, $p_{2}$, $\gamma_{\rm min}$, $\gamma_{\rm max}$,
  $\gamma_{\rm b}$, $N_{0}$ {and} $U_{\rm ext}$ (energy density of external photon fields of BLR or IR). In order to reduce the number of free parameters, the radius of the emitting region in jet frame can be constrained from the minimum variability timescale and redshift with $R \leqslant\delta\Delta t/(1+z)\sim 1.6 \times 10^{15} \delta$ cm, where X-ray variability timescale of $\sim$ 1 day \citep{2014A&A...567.135A} is adopted. The index of electron spectrum, $p_2$, is derived from the fitting the observational X-ray spectrum with a power-law function ($F_{\nu}\propto \nu^{-\alpha}$, $p_2=1+2\alpha$, e.g., \citealt{2012ApJ...752..157Z,2014ApJ...788..104Z}), where the best-fit values of $p_2$ are 4.12 and 5.07 for data of 2009
  and 2013 respectively. {The typical $\gamma_{\rm min}=40$ (e.g., \citealt{2014ApJ...788..104Z}; \citealt{2014ApJS..215....5K}) and
  $\gamma_{\rm max}=1\times10^{8}$ ($\gamma_{\rm max}>>100\gamma_{\rm b}$) are adopted in our fitting, which will not affect our main results.
  The other parameters $B$, $\delta$, $p_{1}$, $\gamma_{\rm b}$, $N_{0}$ or/and $U_{\rm ext}$ keep free in our fitting.}

  The multi-wavelength SEDs of PKS 1424+240 are reproduced using the least-square ($\chi^{2}$) fitting technique (e.g. \citealt{2011ApJ...733...14M};
  \citealt{2012ApJ...752..157Z,2014ApJ...788..104Z}; \citealt{2014ApJS..215....5K}). Similar to \cite{2014A&A...567.135A}, the non-simultaneous
  radio data {(the blue filled square in Figure 1)} is also included in $\chi^{2}$ fitting, due to a possible correlation between radio and optical emission which suggest a common origin of the two bands emission (\citealt{2014A&A...567.135A}). There are 29 observational {data points} (28 simultaneous UV, X-ray and $\gamma$-ray data) in the high state (2009) and 28 observational {data points} (27 simultaneous UV, X-ray and $\gamma$-ray data) in the low state (2013) in the SED modeling. In $\chi^{2}$ fitting, we consider the observational error of the {data points} in the radio, X-ray and $\gamma$-ray band.
  For the UV data with no reported uncertainties, we take $2\%$ of the observational flux as the error (e.g., \citealt{2012ApJ...752..157Z};
  \citealt{2014A&A...567.135A}). We generate all the parameters in a broad range, and calculate the reduced $\chi^{2}_{\rm r}$ for these parameters.
  Then we derive the probability distribution of $\chi^{2}_{\rm r}$ (e.g., $p\propto\exp(-\chi_{\rm r}^{2})$), and the maximum probability corresponds
  the best-fit parameters. The $1\sigma$ uncertainty of each parameter is derived from the Gaussian fit to its probability distribution
   by setting other parameter to its best-fit value (e.g., \citealt{2012ApJ...752..157Z,2014ApJ...788..104Z}; \citealt{2014ApJS..215....5K}).

  The best fits are shown in Figure 1. The dotted, dashed, dot-dashed and solid lines represent the synchrotron, SSC, EC and
  total emission respectively. The left column and right column show the SEDs of high state in 2009 and low state in 2013 respectively, where
  the upper, middle and lower panels correspond to the fitting result using different models (EC-BLR, EC-IR and pure SSC).
  The best-fit parameters, uncertainties and the values of $\chi^2$ are listed in Table \ref{tab:pks1424p240}. We find that the SEDs of the high state (2009) and the low state (2013) can roughly be reproduced by the leptonic jet model with pure SSC, EC-IR and EC-BLR models respectively. Similar to former works (e.g., \citealt{2008ApJ...684L..73A}; \citealt{2011ApJ...726...43A}; \citealt{2012A&A...542A.100A}; \citealt{2012AIPC.1505..647P}; \citealt{2013arXiv1312.4829L}; \citealt{2014ApJ...783...83L}), the quite weak magnetic field and extraordinarily large Doppler factor are needed in the pure SSC model, where $B=0.02$ G, $\delta=51.47$ in low state of 2013 and $B=0.05$ G, $\delta=40.23$ in high state of 2009.
  After considering possible EC component (either IR or BLR seed photons), we find that the model parameters are quite match with those of other
  BL Lacs, where $B=0.1-0.3$ G and $\delta=25-30$ in both states. We note that the energy density of BLR or IR in our modeling is much lower than
  that of luminous FSRQs, where $U_{\rm ext,BLR}=(3.71\pm0.65)\times10^{-5}~{\rm erg~cm}^{-3}$ (high state of 2009),
  $U_{\rm ext,BLR}=(1.62\pm0.48)\times10^{-5}~{\rm erg~cm}^{-3}$ (low state of 2013) and $U_{\rm ext,IR}=(8.24\pm0.33)\times10^{-7}~{\rm erg~cm}^{-3}$
  (high state of 2009), $U_{\rm ext,IR}=(4.09\pm0.37)\times10^{-7}~{\rm erg~cm}^{-3}$ (low state of 2013) in rest frame.

\begin{table*}
\centering
\caption{The relevant parameters of PKS 1424+240 (input model parameters and output luminosities).}\label{tab:pks1424p240}
\begin{tabular}{lcccc ccc}
\hline \hline
Model                                       &\multicolumn{3}{c}{{high state(2009)}}   & &\multicolumn{3}{c}{low state(2013)}         \\
\cline{2-4} \cline{6-8}
Parameter                                       &  (BLR)       &   (IR)      &  (SSC)    & &	(BLR)      &   (IR)         &  (SSC)    \\
\hline
$B$ (G)  	                                    &0.26$\pm$0.11    &0.20$\pm$0.04  &(0.52$\pm$0.04)e-1  &&0.20$\pm$0.03    &0.14$\pm$0.04&(0.22$\pm$0.02)e-1 \\
$\delta$  	                                    &27.28$\pm$2.32   &28.51$\pm$2.67   &40.23$\pm$2.71  &&29.81$\pm$2.89    &31.58$\pm$2.53	&51.47$\pm$4.47 \\
$p_{1}$  	                                    &1.91$\pm$0.03    &1.92$\pm$0.04   &1.93$\pm$0.03     &&1.90$\pm$0.04    &1.91$\pm$0.02   &1.90$\pm$0.05 \\
$\gamma_{\rm b}(10^{4})$    	                &2.45$\pm$0.37    &2.71$\pm$0.43    &4.52$\pm$0.33    &&2.11$\pm$0.46    &2.44$\pm$0.29	    &4.80$\pm$0.42 \\
$N_{0}({\rm cm}^{-3})$                	        &80.96$\pm$31.74  &107.77$\pm$34.55  &99.87$\pm$14.63  &&69.91$\pm$21.25  &88.51$\pm$19.69	&49.41$\pm$9.27 \\
$U_{\rm ext}(10^{-5}{\rm erg~cm}^{-3})$         &3.71$\pm$0.65    &(8.24$\pm$0.33)e-2  &...            &&1.62$\pm$0.48   &(4.09$\pm$0.37)e-2	&...      \\
$\chi^2_{\rm r}$  	                            &1.84             &1.21                 &1.22            &&2.58             &2.36    &2.77     \\
$L_{B}(\rm erg~s^{-1})$  &$9.42\times10^{44}$   &$6.76\times10^{44}$  &$1.74\times10^{44}$  &&$7.93\times10^{44}$  &$4.97\times10^{44}$  &$8.80\times10^{44}$ \\
$L_{\rm e}(\rm erg~s^{-1})$&$4.35\times10^{44}$ &$6.19\times10^{44}$ &$2.22\times10^{45}$  &&$5.12\times10^{44}$  &$7.86\times10^{44}$  &$3.63\times10^{45}$ \\
$\epsilon_{B}=L_{B}/L_{\rm e}$  	            &2.16              &1.09             &0.078          &&1.55                 &0.63                 &0.024 \\
\hline \hline
\end{tabular}
\end{table*}

\section{Conclusion and Discussion}{\label{sec4}}

  In this work, we employ a leptonic model with the least-square ($\chi^{2}$) fitting technique to reproduce the multi-wavelength SED of PKS 1424+240
  in both high (2009) and low (2013) states. Both pure SSC and SSC+EC process are considered in our model even this source is a BL Lac object. After
  including a weak EC component, we find that the model parameters are quite reasonable as in most of other BL Lacs while extreme model parameters are
  needed in pure SSC process. We propose that BLR and/or IR torus may become weak (but not fully disappear) and still play a role in external Compton
  scattering of low-power BL Lacs.

  Pure SSC model was widely adopted in fitting the multi-wavelength SED of HBL BL Lacs (e.g., \citealt{1997A&A...320...19M}; \citealt{2004ApJ...601..151K}; \citealt{2014ApJ...788..104Z}), while luminous FSRQs prefer SSC+EC model (e.g., \citealt{1999ApJ...515..140S}; \citealt{2011ApJ...735..108C}; \citealt{2014MNRAS.439.2933Y}).
  \cite{2002ApJ...581..127B} proposed that the EC from BLR may play a role in some intermediate and low-energy-peaked BL lacs. For HSP PKS 1424+240, in our modeling, both pure SSC and SSC+EC process are considered to model the multi-wavelength SED in both high (2009) and low (2013) states. Nonetheless, the extreme parameters are needed if the pure SSC model is adopted for PKS 1424+240, where the magnetic field is quite low and the Doppler factor is extraordinarily large. This phenomenon is consistent with several other BL Lacs that pure SSC normally need
  extreme input parameters (e.g., \citealt{2008ApJ...684L..73A}; \citealt{2011ApJ...726...43A}; \citealt{2011ApJ...730..101A}; \citealt{2012A&A...542A.100A}; \citealt{2012ApJ...755..118A}; \citealt{2012AIPC.1505..647P}; \citealt{2013arXiv1312.4829L}; \citealt{2013A&A...552A..11R}; \citealt{2014ApJ...783...83L}). After including a weak external photon field, both the magnetic field intensity and Doppler factor are roughly consistent with most of other BL Lacs, where the average values of magnetic field strength $B\sim$ 0.3 G and Doppler factor $\delta\sim$~{10-20}~(see \citealt{2012ApJ...752..157Z}; \citealt{2014Natur.515..376G}). After considering a weak EC component, the ratio of magnetic energy and emitting-electron energy in the blob $\epsilon_{B}=L_{B}/L_{\rm e}\sim 1$ is also quite consistent with other BL Lacs \citep[e.g.,][]{2014Natur.515..376G}, while its value is $<0.1$ in SSC model.
   So, the extreme input model parameters of pure SSC process in both states suggest that the EC process may play a role in PKS 1424+240 even we cannot fully exclude that this source do has some kind of extreme physical properties compared other BL Lacs. It should be noted that the two-zone jet model and lepto-hadronic jet model are also adopted to avoid the extreme jet parameter in this source (\citealt{2014A&A...567.135A},\citealt{2015MNRAS.447.2810Y}), and both of them can give a reasonable fit without the extreme model parameters. Future multi-wavelength variations and better multi-wavelength SED may help to further test this issue.

   The input energy density of external photon fields $U_{\rm BLR}\sim(1-4)\times10^{-5}~{\rm erg~cm}^{-3}$ or $U_{\rm IR}\sim(4-8)\times10^{-7}~{\rm erg~cm}^{-3}$ in low and high state are about 3 orders of magnitude lower than that in luminous FSRQs,
   where $U_{\rm BLR}\sim2.6\times10^{-2}~{\rm erg~cm}^{-3}$ (see \citealt{2008MNRAS.387.1669G}; \citealt{2009MNRAS.397..985G}, for details) and $U_{\rm IR}\sim3\times10^{-4}~{\rm erg~cm}^{-3}$ (see \citealt{2008MNRAS.387.1669G}; \citealt{{2007ApJ...660..117C}}, for details). It should be noted that the required low external seed photon field in PKS 1424+240 is also roughly consistent
   with those of several other BL Lacs (e.g., \citealt{2008ApJ...684L..73A}; \citealt{2011ApJ...726...43A}; \citealt{2011ApJ...730..101A}; \citealt{2012ApJ...755..118A}; \citealt{2012AIPC.1505..647P}; \citealt{2013arXiv1312.4829L}; \citealt{2014ApJ...783...83L}). The physical reason for
   this low seed photon field in BL Lac is not very clear, which may be caused by the weaker BLR or torus still exist in BL Lacs. Normally,
   BL Lac show no or very weak emission-line features due to the possible disappearance of BLR. However, some quite weak broad optical emission
   lines (e.g., $\rm H\alpha$, $\rm Ly\alpha$) are detected in a few BL Lacs (e.g., \citealt{2000MNRAS.311..485C}; \citealt{2012MNRAS.424..393F}; \citealt{2014ApJ...795...57F}). In nearby low-luminosity AGNs, it was also found that broad emission lines and torus are still exist but they are much weaker than luminous Seyfert and QSOs \citep[e.g.,]{2001ApJ...554L..19T,2002ApJ...579..205G,2003ApJ...590...86L,2009ApJ...701L..91E,2010ApJ...724..855C,2007MNRAS.380.1172H,2008ARA&A..46..475H}.
   The weaker BLR or torus in low-luminosity AGNs or low-power BL Lacs may be caused by (or coevolved with) the transition standard cold
   disk to radiatively inefficient accretion flows when accretion rate decreases (e.g., \citealt{2009ApJ...694L.107X}; \citealt{2001A&A...379L...1G}; \citealt{2002ApJ...579..554W}; \citealt{2003ApJ...599..147C}; \citealt{2006PASP..118.1098W}). Our results, combined with several former SED modeling
   of BL Lacs with weak EC components, suggest that BLR or torus may still play a role in Compton scattering in low-power BL Lacs, where BLR/torus may
   become weak but not fully disappear in these low-power sources (e.g., LLAGNs, FR Is and BL Lacs).
   Therefore, the seed photon field density (or BLR/torus) may be strongly evolved with possible change of accretion modes from FSRQs to BL Lacs, where the FSRQs and BL Lacs represent the earlier merger-driven phase (gas-rich, efficient accreting) and final phase (gas-starved, inefficiently accreting) respectively (e.g., \citealt{2014ApJ...780...73A}).

\section*{Acknowledgments}
  We thank the anonymous referee for very constructive and helpful comments and suggestions, which greatly help us to improve our paper. This work is supported by the NSFC (grants 11573009, 11143001, 11103003, 11133005, 11233006, 11463007 and 51567017), the New Century Excellent Talents in University (NCET-13-0238), the Research Foundation for Advanced Talents of Liupanshui Normal University (LPSSYKYJJ201506, LPSSY201401), the Physical Electronic Key Discipline of Guizhou Province (ZDXK201535), the Natural Science Foundation of the Department of Education of Guizhou Province (QJHKYZ[2015]455) and the Science and Technology Foundation of Guizhou Province.

\bsp
\label{lastpage}
\end{document}